\documentclass[
     final            % use final for the camera ready runs
  ]
  {aipproc}

\layoutstyle{6x9}

\usepackage{array}

\newcommand{\kev}{\ensuremath{\mathrm{keV}}}
\newcommand{\mev}{\ensuremath{\mathrm{MeV}}}
\newcommand{\gev}{\ensuremath{\mathrm{GeV}}}
\newcommand{\pipi}{\ensuremath{\pi^+\pi^-}}
\newcommand{\br}{\ensuremath{\mathcal{B}}}

\newcommand{\dz}{\ensuremath{D^0}}
\newcommand{\dzbar}{\ensuremath{\overline{D}{}^0}}
\newcommand{\dstarzb}{\ensuremath{\overline{D}{}^{*0}}}
\newcommand{\dplus}{\ensuremath{D^+}}
\newcommand{\dstarm}{\ensuremath{D^{*-}}}

\newcommand{\ket}[1]{\ensuremath{| #1 \rangle}}

\begin{document}

\title{Charm physics at Belle}

%% PACS codes:
%%
%% 13.66.Bc = ``Specific reactions and phenomenology''
%%            ->''Lepton-lepton interactions''
%%            ->''hadron production in e-e+ interactions''
%%
%% 14.20.Lq = ``Properties of specific particles''
%%            ->''Baryons''->''Charmed baryons''
%%
%% 14.40.Lb = ->''Mesons''->''Charmed mesons''
%%
%% 14.40.Gx = ->''Mesons''->''Mesons with S=C=B=0, mass > 2.5 GeV (incl. quarkonia)''


\author{Bruce Yabsley}{
  address={Virginia Polytechnic Institute and State University, Blacksburg VA 24061},
  altaddress={(for the Belle Collaboration)}
}

\begin{abstract}
  This talk reviews an unrepresentative selection of Belle's
  open-charm and charmonium analyses, focussing on new developments
  and topics of interest to the DIS
  community. Highlights include an $X(3872)$ analysis
  favoring $J^{PC}=1^{++}$, and the $D^0
  \overline{D}{}^{*0}$ bound-state interpretation.
\end{abstract}

\maketitle

%%%%%%%%%%%%%%%%%%%%%%%%%%%%%%%%%%%%%%%%%%%%
%% MAINMATTER
%%%%%%%%%%%%%%%%%%%%%%%%%%%%%%%%%%%%%%%%%%%%

%% INTRODUCTION %%%%%%%%%%%%%%%%%%%%%%%%%%%%

\section{Introduction}

A talk of this length does not allow even a representative survey of
open-charm and charmonium analyses at Belle, so I've made a selection
favoring the most interesting recent developments---concerning the
exotic $X(3872)$ state---and topics of interest to the deep inelastic
scattering community. Due to length limitations, the writeup is even
more cursory than the talk. Interested readers should consult the
references.

The aim of the Belle collaboration is to study violation of
the CP symmetry, using the time-dependence of decays of
$B\overline{B}$ pairs.  Open-charm and charmonium studies are an
active sideline. The KEKB collider~\cite{kekb} produces $e^+e^- \to
\Upsilon(4S) \to B\overline{B}$ and $e^+e^- \to q\bar{q}$ continuum
events with unprecedented luminosity: both $B$-decays and the
continuum are copious sources of charmed and charmonium states. The
Belle detector~\cite{belle}, at the KEKB interaction point, is a
general-purpose detector with good particle ID %identification
capabilities.

%% X(3872) %%%%%%%%%%%%%%%%%%%%%%%%%%%%%%%%%

\section{The $X(3872)$: quantum numbers and interpretation}

\begin{figure}
  \begin{tabular}{c@{\hspace{0.05\textwidth}}c}
    \raisebox{3ex}{\includegraphics[width=0.45\textwidth]{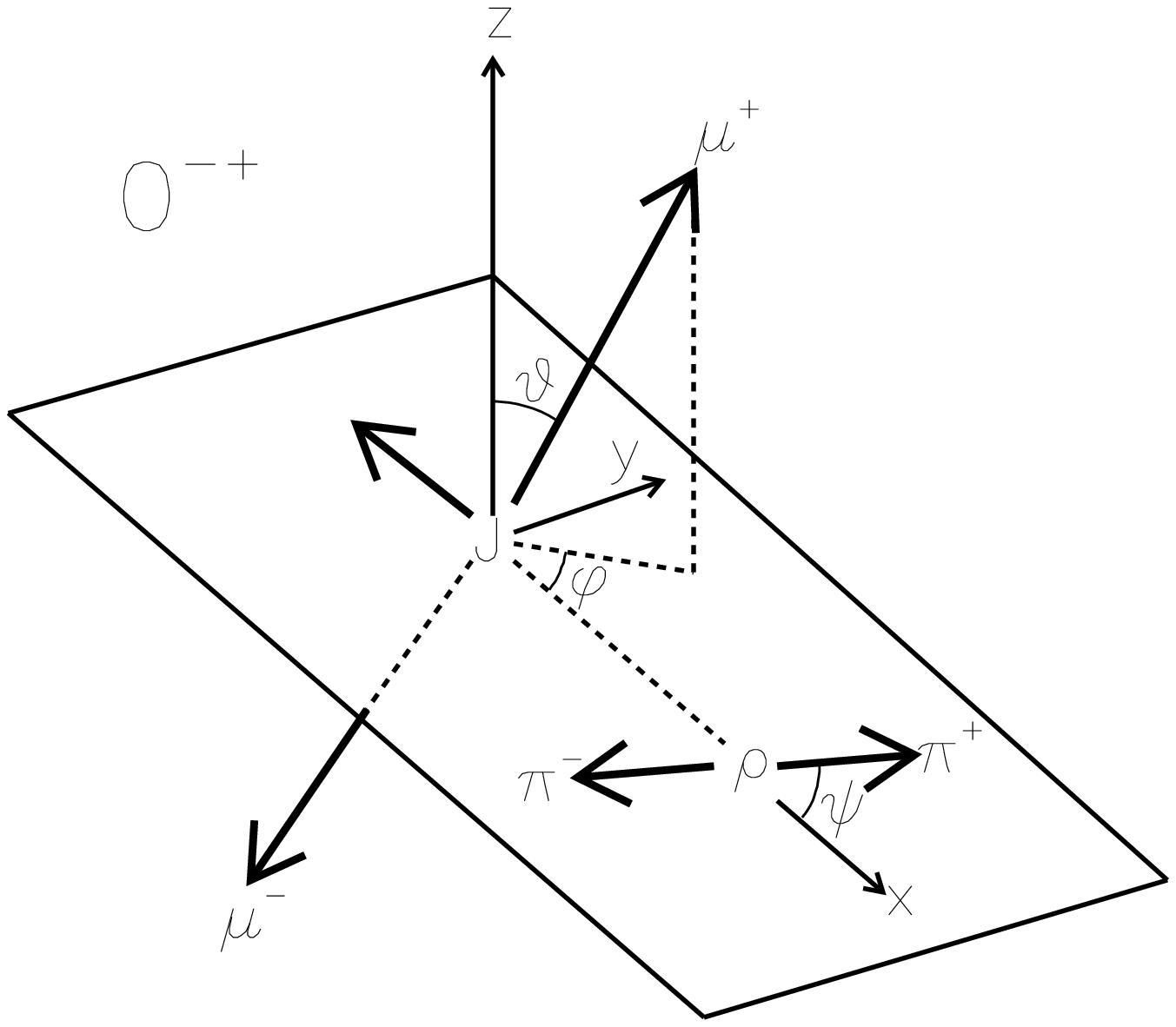}}
  & \includegraphics[width=0.45\textwidth]{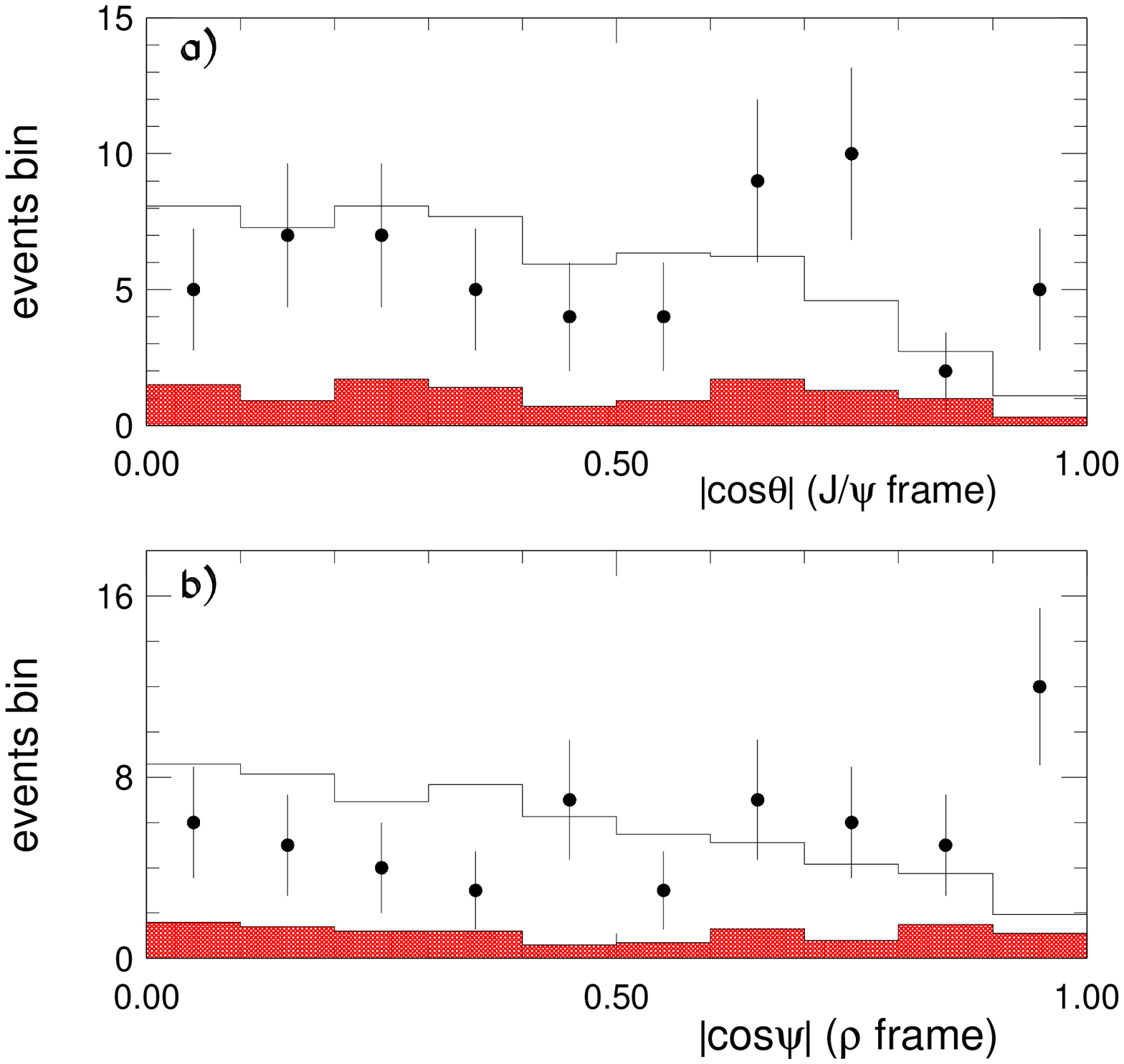}
  \end{tabular}
  \caption{$X(3872) \to \pipi\,J/\psi$ angular distributions for data (points),
    and for the $J^{PC} = 0^{-+}$ hypothesis (histogram), 
    including background estimated from $X$-mass sidebands (shaded). 
    The definition of the angles is shown in the sketch on the left.
    The $\chi^2$ of the fits are (a) $17.7$ and (b) $34.2$ for 9 degrees of freedom,
    disfavoring $0^{-+}$. Note the concentration of events in the final bins,
    contrary to expectation.}
  \label{fig-X3872-0+-}
\end{figure}

The $X(3872)$, a narrow state decaying to $\pipi\,J/\psi$, was
discovered in $B \to K \pipi\,J/\psi$ decays by
Belle~\cite{x3872-belle-discovery}, and confirmed by three other
groups~\cite{x3872-confirmation}. In subsequent analysis, it has not
been possible to match the properties of the $X$ with
those of an expected $c\bar{c}$ state~\cite{x3872-belle-charmonium}.  Belle has recently reported
the observation of $X(3872)\to\gamma\,J/\psi$ and $\omega\,J/\psi$
decays~\cite{x3872-belle-gpsi-omegapsi}, confirming that the C-parity
of the $X$ must be even.  An angular analysis of $X(3872)$
decays has also been performed~\cite{x3872-belle-jpc}, exploting
zeroes in predicted distributions~\cite{x3872-rosner} to
test various $J^{PC}$ hypotheses. An example is shown in
Fig.~\ref{fig-X3872-0+-}.

\begin{figure}
%  \begin{center}
    \includegraphics[width=0.60\textwidth,bb=84 240 509 500]
		    {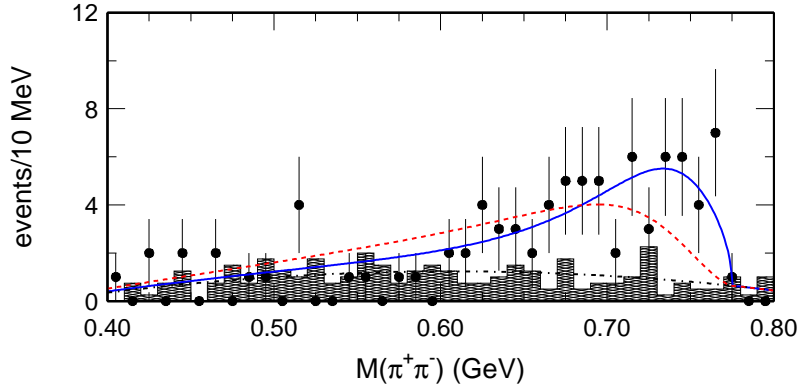}
%  \end{center}
  \caption{$M(\pi^+\pi^-)$ distribution for events in the $X(3872)$ signal region (points)
    and sideband (shaded). Fits to $J^{++}$ (solid) and $J^{+-}$ (dashed) hypotheses
    are also shown: see the text.}
  \label{fig-X3872-mpipi}
\end{figure}

The $X \to \pipi\,J/\psi$ dipion mass distribution 
is shown in Fig.~\ref{fig-X3872-mpipi}. The rate
near the kinematic boundary is sensitive to the parity of the $X$: for
$C=+1$ and even parity, $q_{J/\psi}^*$ dependence is expected
($\rho$ and $J/\psi$ in S-wave, ignoring D-wave admixture); for
odd parity, $(q_{J/\psi}^*)^3$ (P-wave; ignoring F-wave). 
Fits to the two cases find
$\chi^2 = 43.1$ and $71.0$, for 39 degrees of freedom,
favoring $J^{++}$ hypotheses.
$J^{PC} = 0^{++}$ is disfavored by angular distributions, and 
$J^{PC} = 2^{++}$ by preliminary evidence for decays to
$\dz\dzbar\pi^0$~\cite{x3872-belle-ddbarpi}.
 
The $J^{PC} = 1^{++}$ hypothesis is consistent with available data; 
all other assignments are disfavored by at least one test. However,
the identification of the $X(3872)$ with the $1^{++}$ charmonium state
$\chi_{c1}^\prime$ is unlikely: potential model predictions
for the $\chi_{c1}^\prime$ mass are 3953--$3990\,\mev$, and shift
upward when coupling to open charm 
is taken into account~\cite{x3872-chic1p}.
The isospin-violating $\chi_{c1}^\prime\to\pipi J/\psi$ decay would
presumably have a small partial width, similar to
$\Gamma(\psi(2S)\to\pi^0 J/\psi) = (0.27 \pm 0.06)\,\kev$~\cite{pdg},
to be compared with a total width $\Gamma >
1\,\mev$~\cite{x3872-chic1p}.
This contradicts BaBar's 90\% confidence limit 
$\br(X(3872)\to\pipi\,J/\psi) > 4.3\%$~\cite{x3872-babar-lower}.
The low ratio of radiative and hadronic partial
widths $\Gamma(X\to\gamma\psi)/\Gamma(X\to\pipi\psi) = 0.14 \pm
0.05$~\cite{x3872-belle-gpsi-omegapsi} also disfavors
$\chi_{c1}^\prime$.

By contrast, the observed properties of the $X(3872)$ are consistent
with those of a $\dz\dstarzb$ bound state~\cite{x3872-swanson}:
the mass is within errors of \dz\dstarzb\ threshold, $(+0.6 \pm 1.1)\,\mev$;
as the mass difference $M(\dplus\dstarm) - M(\dz\dstarzb) =
8.1\,\mev$ is large by comparison, isospin violation is natural for
such a state, explaining the observation of 
$X(3872) \to \omega J/\psi$ \emph{and} $\rho J/\psi$ decays. These
decays are natural within the model of
Swanson~\cite{x3872-swanson}, where 
$\ket{\omega J/\psi}$ and 
$\ket{\rho J/\psi}$ appear as admixtures to the $\ket{\dz\dstarzb}$
wavefunction. A small branching ratio
$\Gamma(X\to\gamma J/\psi)/\Gamma(X\to\pipi J/\psi)$ is also expected
for such a state.

%% Y(3940) %%%%%%%%%%%%%%%%%%%%%%%%%%%%%%%%%
%
%\section{The ``$Y(3940)$'' enhancement in $M(\omega J/\psi)$}
%
% (moved to spectroscopy, below)

%% 2ccbar production %%%%%%%%%%%%%%%%%%%%%%%

\section{Double charmonium production and the $X(3940)$}

The process $e^+e^- \to c\bar{c} c\bar{c}$ was discovered by Belle in
both double charmonium ($J/\psi\,\eta_c$) and associated charm
($J/\psi\,D^{(*)}\,X$) production; both processes have unexpectedly
large rates~\cite{cccc-belle-discovery}. Various proposed alternative
explanations of the data have been contradicted by further tests, 
including angular analysis and full reconstruction of $e^+e^-
\to J/\psi\,\eta_c$ events~\cite{cccc-belle-detail}. The principal
results have recently been confirmed by BaBar~\cite{cccc-babar}.

Evidence for a new state, $X(3940)$, seen in the recoil mass ($M_X$)
spectrum in $e^+e^- \to J/\psi\,X$ events, was presented in 2004 at
the Beijing conference~\cite{x3940-2004}. Decays of this state favor
$D\overline{D}{}^*$, based on a study of events with reconstructed
$J/\psi$ and $D$ mesons. An updated analysis confirming these results
is being prepared for publication this summer~\cite{x3940-2005}.

%% other spectroscopy %%%%%%%%%%%%%%%%%%%%%%

\section{Other results (mostly spectroscopy)}

An enhancement at $\omega\,J/\psi$ threshold has
been seen in $B\to K\omega\,J/\psi$~\cite{cccc-belle-discovery}. Interpreted as a particle
($M=(3943\pm11\pm13)\,\mev$, $\Gamma=(87\pm22\pm26)\,\mev$), this ``$Y(3940)$'' would
be exotic: a $c\bar{c}$ state at this mass would be expected to
decay to $D\overline{D}{}^{(*)}$, with very small branching
fractions for $\omega\,J/\psi$ and other hadronic charmonium transitions.

%\subsection{The $D_{sJ}$ states}
Belle observed the $D_{sJ}^*(2317)$ and $D_{sJ}(2460)$
in both continuum production~\cite{dsj-belle-continuum}
and $B$ decays~\cite{dsj-belle-bdecay},
confirming the observations by BaBar and CLEO~\cite{dsj-babar,dsj-cleo},
establishing the $D_{sJ}(2460)^+ \to \gamma D_s^+$ decay, 
and favoring $J^P(D_{sJ}(2460)) = 1^+$, based on the 
$\gamma D_s^+$ helicity angle distribution. 
Study of $D_{sJ}^*(2317)^+ \to \pi^0 D_s^+$~\cite{dsj-belle-bdecay-extended}
likewise favors $J^P = 0^+$.

%\subsection{Search for the SELEX state}
Searches for the $D_{sJ}(2632)^+ \to D_s^+ \eta$ and $D^0 K^+$ state
of SELEX~\cite{dsj-selex} find no evidence of production in
$B$ decays or the continuum at Belle~\cite{dsj-selex-belle}.

%\subsection{Observation of the $\Sigma_c(2800)$}
Amongst other charmed baryon results, a new isospin triplet
$\Sigma_c(2800)$ decaying to $\Lambda_c^+ \pi^{-,0,+}$ has been
observed in the continuum~\cite{sigmac2880}. It is tentatively
identified as the $\Sigma_{c2}$ ($J^P = 3/2$), with some admixture of
the $\Sigma_{c1}$ (with the same quantum numbers).

%\subsection{Search for the $\Theta(1540)$ pentaquark}
Of the Belle pentaquark searches reported in
2004~\cite{pq-belle-2004}, the most important uses interactions of
kaons (from $e^+e^-$ annihilation) with the material of the
detector. This study placed a limit on production of the $\Theta(1540)$
relative to the $\Lambda(1520)$: an updated analysis, to be published
in the summer of 2005~\cite{pq-belle-2005},
also bounds the
rate of exclusive production $K^+ n \to \Theta(1540)^+ \to p K^0_S$,
with similar sensitivity to that of DIANA~\cite{pq-diana}.

%% charm fragmentation %%%%%%%%%%%%%%%%%%%%%

% \section{Charm fragmentation}

A major study of charm fragmentation in $e^+e^- \to c\bar{c}$ at
$\sqrt{s} \simeq 10.6\,\gev$ will also be submitted for publication
this summer~\cite{fragmentation}. Fractional momentum distributions for $D$, $D_s$, and $D^*$
mesons, and the $\Lambda_c^+$, are measured with much
greater precision than in previous studies; a comparison of
fragmentation functions is also presented.

%% summary %%%%%%%%%%%%%%%%%%%%%%%%%%%%%%%%%

\section{Summary}

Recent Belle analyses include a study of $X(3872)$
% NOTE: the commented-out line (from arXiv v1) contains a typo
%decays and properties, favouring $J^{PC} = 1^{++}$ and the $\dz\dzbar$
decays and properties, favouring $J^{PC} = 1^{++}$ and the $\dz\dstarzb$
molecular model. Other contributions to spectroscopy include the
$Y(3940)$, double charmonium production (including $e^+e^-
\to J/\psi\,X(3940)$), and pentaquark searches. A study of charm
fragmentation has also been performed. Many other results in charm and
charmonium studies lie outside the scope of this talk.

%%%%%%%%%%%%%%%%%%%%%%%%%%%%%%%%%%%%%%%%%%%%%%%%
%% BACKMATTER
%%%%%%%%%%%%%%%%%%%%%%%%%%%%%%%%%%%%%%%%%%%%%%%%

% tweak to duplicate proceedings 4 pages on arXiv LaTeX installation
\vspace*{-1.0ex}

\begin{theacknowledgments}
  I'd like to thank my colleagues at Belle for their help in preparing
  this talk, and the organisers of the Deep Inelastic Scattering
  workshop. The work of Madison's fine local breweries was also
  greatly appreciated. 
\end{theacknowledgments}

% tweak to duplicate proceedings 4 pages on arXiv LaTeX installation
\vspace*{-1.0ex}

\bibliographystyle{aipproc}   % if natbib is available

\end{document}